\documentclass[dvipdfmx,11pt,a4paper]{article}
\usepackage{amsmath}
\usepackage{amssymb}
\usepackage{ascmac}
\usepackage{amsthm}
\usepackage{slashed}
\usepackage{bm}
\usepackage[dvipdfmx]{graphicx}
\usepackage[centering,headsep=15pt,head=80pt,foot=20pt,margin=60pt]{geometry}
\usepackage[scaled]{helvet}
\usepackage{color}

\newcommand{\Sla}[1]%
{\kern0.12em{\raise.15ex\hbox{$/$}\kern-.74em #1}}

\renewcommand{\Re}{\mathop{\rm Re}}
\renewcommand{\Im}{\mathop{\rm Im}}

\newcommand{\Group}[2]{{ \hbox{{\itshape{#1}}($#2$)} }}
\newcommand{\U}[1]{\Group{U\kern0.05em}{#1}}
\newcommand{\SU}[1]{\Group{SU\kern0.1em}{#1}}
\newcommand{\SL}[1]{\Group{SL\kern0.05em}{#1}}
\newcommand{\Sp}[1]{\Group{Sp\kern0.05em}{#1}}
\newcommand{\SO}[1]{\Group{SO\kern0.1em}{#1}}

\newcommand{\mybar}[1]%
    {{\kern 0.8pt\overline{\kern -0.8pt#1\kern -0.8pt}\kern 0.8pt}}
\newcommand{\sla}[1]%
    {{\raise.15ex\hbox{$/$}\kern-.57em #1}}
\newcommand{\roughly}[1]%
    {{ \mathrel{\raise.3ex\hbox{ $#1$\kern-.75em\lower1ex\hbox{$\sim$}} } }}

\newcommand{\nop}[1]{:\kern-.3em#1\kern-.3em:}

\newcommand{\al}{\ensuremath{\alpha}}
\newcommand{\be}{\ensuremath{\beta}}

\newcommand{\de}{\ensuremath{\delta}}

\newcommand{\ep}{\ensuremath{\epsilon}}

\renewcommand{\th}{\ensuremath{\theta}}

\newcommand{\la}{\ensuremath{\lambda}}
\newcommand{\La}{\ensuremath{\Lambda}}

\newcommand{\ph}{\ensuremath{\phi}}

\newcommand{\ch}{\ensuremath{\chi}}

\newcommand{\GeV}{ \ensuremath{\mathrm{\,GeV}} }
\newcommand{\TeV}{ \ensuremath{\mathrm{\,TeV}} }

\newcommand{ \pb}{ \ensuremath{\mathrm{ \,pb}} }

\newcommand{\invfb}{ \ensuremath{\mathrm{ \,fb^{-1}}} }

\newcommand{\mcl}[1]{\mathcal{#1}}

\numberwithin{equation}{section}
\numberwithin{figure}{section}

\begin{document}

\begin{titlepage}

\begin{flushright}
UT-14-11 \\
UG-FT-309/14\\
CAFPE-179/14\\
\end{flushright}

\vskip 5em

\begin{center}
{\Large \bfseries
 Unitarity Bounds on Dark Matter Effective Interactions at LHC
}

\vskip 4em

Motoi Endo$^\sharp$,
Yasuhiro Yamamoto$^\flat$

\vskip 4em

$^\sharp$ 
  \textit{Department of Physics, Faculty of Science, The University of Tokyo, }\\
  \textit{Tokyo 133-0022, Japan}\\[1em]
$^\flat$
  \textit{Deportamento de Fisica Teorica y del Cosmos,}\\
  \textit{Facultad de Ciencias, Universidad de Granada,}\\
  \textit{Granada E-18071 Spain}\\

\vskip 4em

\textbf{Abstract}
\end{center}

\medskip
\noindent

The perturbative unitarity bound is studied in the monojet process at LHC.
The production of the dark matter is described by the low-energy effective theory. 
The analysis of the dark matter signal is not validated, if the unitarity condition is violated. 
It is shown that the current LHC analysis with the effective theory breaks down, at least, when the dark matter is heavier than $\mcl{O}(100)\GeV$.
Future prospects for $\sqrt{s}=14\TeV$ are also discussed.
The result is independent of physics in higher energy scales. 

\bigskip
\vfill
\end{titlepage}

\section{Introduction}

The existence of the dark matter is an evidence of physics beyond the standard model (SM). 
A lot of efforts have been devoted to search for it. 
In collider experiments, the dark matter is expected to be detected as a missing component, or momentum unbalance of visible particles. 
At LHC, monojet signatures with a large missing transverse energy (MET) have been studied~\cite{ATLAS:2012zim,CMS:rwa}.\footnote
{
Monojet signatures at hadron colliders have been discussed in Ref.~\cite{Beltran:2010ww}.
Those at Tevatron are in Ref.~\cite{Bai:2010hh}.
}
Since no excess has been measured over backgrounds, production cross sections of the dark matter are constrained. 

In the study of the monojet signature, the low-energy effective theory of the dark matter is considered~\cite{Goodman:2010ku}. 
Interactions with the SM particles are supposed to be described by higher dimensional operators. 
Such an approach is superior, since the result is independent of details of physics in ultraviolet (UV) energy scale. 
However, the analysis is not validated, once the effective theory breaks down. 
This happens when events do not satisfy the perturbative unitarity bound of the effective theory. 
In this letter, we study the unitarity bound in the LHC process, $pp \to \chi\bar\chi j$, where $\chi$ is the dark matter. 
It will be shown that the LHC analysis based on the effective theory is not guaranteed when the dark matter is heavier than $\mcl{O}(100)\GeV$.
In the region, a number of signal events violate the unitarity bound. 

Before proceeding to the study of the unitarity bound, let us mention previous works. 
In Ref.~\cite{Shoemaker:2011vi}, the unitarity bound on the monojet signature was studied. 
The analysis focused on the vector effective operator (see Sec.~\ref{sec:VV}).
Since the center-of-mass (c.m.) energy of the subprocess, $q\bar q \to \chi\bar\chi$, is set to be that of the proton beam, the result looks too strong. 
We take the parton distribution functions (PDFs) into account. 
In Refs.~\cite{Busoni:2013lha,Buchmueller:2013dya}, the validity of the effective theory of the dark matter was studied by assuming s-channel mediators for the UV physics.
Their result relies on the mediator, whereas the unitarity bound in this letter is independent of the UV completion. 

This letter is organized as follows. 
In Sec.~\ref{SecUnitarity}, the unitarity bound is reviewed. 
The bound is shown in terms of the helicity amplitudes. 
In Sec.~\ref{sec:lagrangian}, the effective theory of the dark matter which we consider is summarized.  
The helicity amplitudes and the unitarity bound are listed for each operator.
In Sec.~\ref{sec:status}, the analysis is applied to the current results of ATLAS and CMS. 
In Sec.~\ref{sec:LHC14TeV}, future prospects of the unitarity bound are studied for LHC at $\sqrt{s} = 14\TeV$.
Cut dependence is also discussed. 
The bound depends on the signal region, especially on the cut threshold of the jet transverse momentum and the missing transverse energy. 
In Sec.~\ref{sec:conclusion}, we summarize the results.

\section{Unitarity Bound}
\label{SecUnitarity}

From the optical theorem, amplitudes of $2\to 2$ scattering processes satisfy the following relation:
\begin{align}
 2\Im[ \mcl{M}_{i\to i}] =\sum_f \be_f \int \frac{\text{d}\cos \th}{16\pi} |\mcl{M}_{i\to f}|^2,
  \label{eq:optical}
\end{align}
where $i$ and $f$ respectively stand for initial and final states, and $\th$ is the angle between directions of the initial and final state particles in the c.m.~frame. 
The factor $\be_f$ is given by the phase space integral as
\begin{align}
 \be_f = \frac{\sqrt{[s-(m_1+m_2)^2][s-(m_1-m_2)^2]}}{s},
\end{align}
where $m_{1,2}$ are masses of the final state particles.
It corresponds to the velocity of the final state particles, if $m_1=m_2$ is satisfied.
The scattering processes are described by the helicity amplitudes~\cite{Jacob:1959at}, where the initial and final state particles are represented by helicity eigenstates. 
Wave functions of spin 1/2 and spin 1 particles are summarized in Appendix~\ref{sec:helicity}.
The amplitude, $\mcl{M}_{i\to f}$, can be expanded by partial waves in terms of the Wigner $d$-function, $d_{\la_f \la_i}^j (\th)$, as
\begin{align}
  \mcl{M}_{i\to f}(\cos\th) = 8\pi \sum_{j=0}^\infty (2j+1) T_{i\to f}^j d^j_{\la_f \la_i}(\th),
  \label{eq:helicity}
\end{align}
where $\la_i$ ($\la_f$) is the total helicity of the initial (final) state.
Here and hereafter, we consider azimuthally symmetric processes, and $\phi = 0$ is chosen without loss of generality. 
The Wigner $d$-function satisfies the relations,
\begin{align}
 d^j_{\la',\la}(0) = \de_{\la'\la}, \qquad
 \int_{-1}^1 \text{d}\cos \th\, d^j_{\la',\la}(\th) d^{j'}_{\la',\la}(\th)
 =\frac{2}{2j+1}\de_{j'j}.
\end{align}
From Eqs.~\eqref{eq:optical} and \eqref{eq:helicity}, the following equation is obtained for each mode of $j$:
\begin{align}
 2\Im[T_{i\to i}^j] 
 = \sum_f \be_f |T_{i\to f}^j|^2
 = \be_i |T_{i\to i}^j|^2 + \sum_{f \neq i} \be_f |T_{i\to f}^j|^2.
\end{align}
The unitarity bound for an elastic channel becomes
\begin{align}
 \be_i\, \Re[T_{i\to i}^j] \leq 1, \qquad
 \be_i\, \Im[T_{i\to i}^j] \leq 2.
\end{align}
For an inelastic scattering amplitude, $T_{i\to f}^j$, the perturbative unitarity condition is obtained for each $j$ as\footnote{
If initial (final) state particles are identical, the unitarity bound is obtained by replacing 
$\be_i \to \be_i/2$ ($\be_f \to \be_f/2$) for $2\to 2$ scatterings.
}
\begin{align}
 \sum_{f \neq i} \be_i \be_f |T_{i\to f}^j|^2 \leq 1.
\label{EqUnitarity}
\end{align}
Since dark matter productions are considered in the proton collisions, we require that their scattering amplitudes satisfy the unitarity bound \eqref{EqUnitarity}.

\section{Effective Lagrangian}
\label{sec:lagrangian}

In this letter, productions of the dark matter at LHC are studied. 
The dark matter is assumed to be a Dirac fermion. 
It is produced through the scattering of partons with a monojet $j$ as
\begin{align}
  pp \to q\bar q j \to \chi\bar\chi j, \qquad
  pp \to gg j \to \chi\bar\chi j.
  \label{eq:process}
\end{align}
In the second step of each process, the dark matters are produced through the parton-scattering subprocesses, $ii \to \chi\bar\chi$ ($i = q$ or $g$).
We consider that the subprocesses are represented by the effective theory. 
The following higher dimensional operators are studied: the scalar, pseudo-scalar, vector, axial-vector and gluon interactions between the dark matter and quarks/gluons.  
Their definitions are given in the subsections. 

The perturbative unitarity bound \eqref{EqUnitarity} is applied to subprocesses of the monojet process \eqref{eq:process}.
Scattering amplitude of the subprocess $ii \to \chi\bar\chi$ is enhanced when $\sqrt{\bar s}$ is comparable to or larger than the suppression scale of the effective operator. 
Here and hereafter, $\sqrt{\bar s}$ denotes the c.m.~energy of the subprocess, i.e., the invariant mass of the dark matters.
Thus, the unitarity bound may be essential when we test the validity of the analysis of the monojet process within the effective theory. 
It is stressed that the bound is independent of the UV physics. 
This cannot be evaded, for instance, by modifying the mediators.

In Eq.~\eqref{EqUnitarity}, $\be_i = 1$ is satisfied, while $\be_f$ for the final state particle (dark matter) becomes
\begin{align}
  \be_f = \sqrt{1-\frac{4m_{\rm DM}^2}{\bar s}},
\end{align}
where $m_{\rm DM}$ is the mass of the dark matter.
In the following, the unitarity bound is individually studied for each effective operator, and interferences among them are discarded.
Some remarks on these procedures are discussed in Sec.~\ref{sec:discussion}.

\subsection{Scalar Operator}
\label{eq:SS}

The scalar effective operator is defined as
\begin{align}
  \mcl{L} = \frac{1}{\La_S^2} (\bar\chi\chi) (\bar q q),
  \label{eq:operatorS}
\end{align}
where $\chi$ is a Dirac dark matter, and $q$ is the up-type or down-type quark. 
It is dimension-6 operator with the suppression scale $\La_S$.
The dark matter production proceeds through the scattering of the quarks. 
In the partial wave expansion, the helicity amplitudes of $q\bar q\to \chi\bar\chi$ are calculated as
\begin{align}
  T^{j=0}_{(+,+)_i\to(+,+)_f} = T^{j=0}_{(+,+)_i\to(-,-)_f} = 
  \frac{1}{8\pi} \frac{\bar s}{\La_S^2} \sqrt{1-\frac{4m_{\rm DM}^2}{\bar s}},
  \label{eq:amplitudeS}
\end{align}
by using the wave functions in the helicity basis (see Appendix~\ref{sec:helicity}).
In the subscriptions, $(\la_a,\la_b)_i$ and $(\la_c,\la_d)_f$ are helicities of the initial and the final state particles in the scattering process $a+b \to c+d$, respectively.
Since they are spin 1/2 states, $\pm$ means $\la = \pm 1/2$.
The particle $b$ ($d$) propagates the opposite direction to the particle $a$ ($c$) in the c.m.~frame. 
Its helicity is defined as $-\la$ in the convention of this letter. 
Thus, the total helicity in the Wigner $d$-function, $d_{\la_f \la_i}^j (\th)$, satisfies $\la_i = \la_a - \la_b$ and $\la_f = \la_c - \la_d$.
The scattering proceeds only when the initial helicity is either $(+,+)_i$ or $(-,-)_i$.
In the above, the initial helicity $(+,+)_i$ is chosen. 
If all the helicities are flipped as $+ \leftrightarrow -$, the amplitudes become the same as above, while the other helicity amplitudes are zero.

In Eq.~\eqref{EqUnitarity}, only the amplitude with $j=0$ provides a non-trivial bound. 
The summation runs over the final state, $f = (+,+)_f, (+,-)_f, (-,+)_f, (-,-)_f$, though amplitudes of two of them vanish according to Eq.~\eqref{eq:amplitudeS}.
The unitarity bound becomes
\begin{align}
  \La_S \geq 
  \left[
    \frac{\bar s^2}{32\pi^2} \left(1-\frac{4m_{\rm DM}^2}{\bar s}\right)^{3/2}
  \right]^{1/4}.
  \label{eq:unitarityS}
\end{align}
If we turn on only the scalar operator \eqref{eq:operatorS} in the effective theory, the monojet process takes place when the initial helicity is either $(+,+)_i$ or $(-,-)_i$. 
All the scatterings are constrained by the above condition. 
The LHC signal events are required to satisfy the condition \eqref{eq:unitarityS} not to ruin the calculation.
Otherwise, the effective approach to the scattering rate is not validated, but rather, the UV physics is needed for the analysis.

\subsection{Psuedo-scalar Operator}
\label{eq:PP}

The pseudo-scalar effective operator is defined with the suppression scale $\La_P$ as
\begin{align}
  \mcl{L} = \frac{1}{\La_P^2} (\bar\chi\gamma_5\chi) (\bar q\gamma_5 q),
\end{align}
where $q$ is quarks, $q = u, d$. 
The situation is similar to that of the scalar operator in Sec.~\ref{eq:SS}.
The helicity amplitudes of $q\bar q\to \chi\bar\chi$ are calculated as
\begin{align}
  T^{j=0}_{(+,+)_i\to(+,+)_f} = -T^{j=0}_{(+,+)_i\to(-,-)_f} = 
  \frac{-1}{8\pi} \frac{\bar s}{\La_P^2},
\end{align}
and the same for $+ \leftrightarrow -$. The other amplitudes vanish. 
The unitarity bound becomes
\begin{align}
  \La_P \geq 
  \left[
    \frac{\bar s^2}{32\pi^2} \sqrt{1-\frac{4m_{\rm DM}^2}{\bar s}}
  \right]^{1/4}.
\end{align}

\subsection{Vector Operator}
\label{sec:VV}

The vector effective operator is defined with the suppression scale $\La_V$ as
\begin{align}
  \mcl{L} = \frac{1}{\La_V^2} (\bar\chi\gamma^\mu\chi) (\bar q\gamma_\mu q),
\end{align}
where $q$ is quarks, $q = u, d$. 
The helicity amplitudes of $q\bar q\to \chi\bar\chi$ become
\begin{align}
  & T^{j=1}_{(+,-)_i\to(+,-)_f} = T^{j=1}_{(+,-)_i\to(-,+)_f} = 
  \frac{-1}{12\pi} \frac{\bar s}{\La_V^2}, \\
  & T^{j=1}_{(+,-)_i\to(+,+)_f} = T^{j=1}_{(+,-)_i\to(-,-)_f} = 
  \frac{-1}{6\sqrt{2}\pi} \frac{m_{\rm DM}\sqrt{\bar s}}{\La_V^2},
\end{align}
and the same for $+ \leftrightarrow -$.
The amplitudes become non-zero when the initial helicity is either $(+,-)_i$ or $(-,+)_i$, because the interaction is vector-like.
The scattering with the final state helicity, $(+,+)_f$ or $(-,-)_f$, takes place by flipping the chirality with the dark matter mass. 
In the partial wave expansion, only the amplitude with $j=1$ is non-zero due to the initial state helicity. 
The amplitude with $(-,+)_i$ is obtained by flipping the above helicities as $+ \leftrightarrow -$.
The other helicity amplitudes vanish.
Since the initial helicity is either $(+,-)_i$ or $(-,+)_i$ to induce the monojet process, the LHC signal events is required to satisfy the unitarity bound,
\begin{align}
  \La_V \geq 
  \left[
    \frac{\bar s(\bar s+2m_{\rm DM}^2)}{72\pi^2} \sqrt{1-\frac{4m_{\rm DM}^2}{\bar s}}
  \right]^{1/4}.
\end{align}

\subsection{Axial-vector Operator}

The axial-vector effective operator is defined with the suppression scale $\La_A$ as
\begin{align}
  \mcl{L} = \frac{1}{\La_A^2} (\bar\chi\gamma^\mu\gamma_5\chi) (\bar q\gamma_\mu\gamma_5 q),
\end{align}
where $q$ is quarks, $q = u, d$. 
The situation is similar to that of the vector operator in Sec.~\ref{sec:VV}.
The helicity amplitudes of $q\bar q\to \chi\bar\chi$ become
\begin{align}
  T^{j=1}_{(+,-)_i\to(+,-)_f} = - T^{j=1}_{(+,-)_i\to(-,+)_f} = 
  \frac{-1}{12\pi} \frac{\bar s}{\La_A^2} \sqrt{1-\frac{4m_{\rm DM}^2}{\bar s}},
\end{align}
and the same for $+ \leftrightarrow -$.
The unitarity bound is
\begin{align}
  \La_A \geq 
  \left[
    \frac{\bar s^2}{72\pi^2} \left(1-\frac{4m_{\rm DM}^2}{\bar s}\right)^{3/2}
  \right]^{1/4}.
\end{align}

\subsection{Gluon Operator}

The gluon effective operator is defined as
\begin{align}
  \mcl{L} = \frac{\al_S}{16\pi\La_K^3} (\bar\ch\ch) G^a{}_{\mu\nu}G^{a\mu\nu},
\end{align}
where $G^a{}_{\mu\nu}$ is the field strength of the gluon, and $\al_S$ is the strong coupling constant. 
It is dimension-7 operator with the suppression scale $\La_K$.
The dark matter is produced in the monojet process through the scattering of gluons. 
By using the wave functions of the massless vector as well as that of the spin 1/2 particle in \ref{sec:helicity}, 
the helicity amplitudes of $gg \to \chi\bar\chi$ are calculated as
\begin{align}
  T^{j=0}_{(+,+)_i\to(+,+)_f} = T^{j=0}_{(+,+)_i\to(-,-)_f} = 
  -\frac{\al_S}{64\pi^2} \frac{\bar s\sqrt{\bar s-4m_{\rm DM}^2}}{\La_K^3},
\end{align}
where $(\la_a,\la_b)_i$ is helicities of the initial state gluons, and $(\la_c,\la_d)_f$ is those of the final state dark matters. 
Since the gluon is a spin 1 particle, $\pm$ in the initial state means $\la = \pm 1$.
As expected from the angular momentum conservation, the helicity amplitudes become non-zero when the total helicity is equal to zero, $j=0$. 
The amplitudes with the opposite helicities, $+ \leftrightarrow -$, become the same as above, whereas the others are zero.
From Eq.~\eqref{EqUnitarity}, the unitarity bound for the gluon operator becomes
\begin{align}
  \La_K \geq 
  \left[
    \frac{\alpha_s^2 \bar s^3}{2048\pi^4} \left(1-\frac{4m_{\rm DM}^2}{\bar s}\right)^{3/2}
  \right]^{1/6}.
\end{align}
Similarly to the previous operators, the monojet process is required to satisfy the condition \eqref{EqUnitarity}, otherwise the effective theory approach is not appropriate to calculate the cross section. 

\section{Present Status}
\label{sec:status}

The monojet signature has been studied at LHC with $\sqrt{s}=8\TeV$~\cite{ATLAS:2012zim,CMS:rwa}.
No excess has been measured over the backgrounds, and the suppression scale $\La$ is constrained from below. 
Since the beam energy is much higher than the limit on $\La$, a part of the signal events may violate the unitarity condition \eqref{EqUnitarity}. 
In this section, we apply the above perturbative unitarity bounds to the current study of the monojet signature at ATLAS and CMS.

\subsection{ATLAS at 8\,TeV}
\label{sec:ATLAS}

\begin{figure}[tb]
\centering
 \includegraphics[width=7cm]{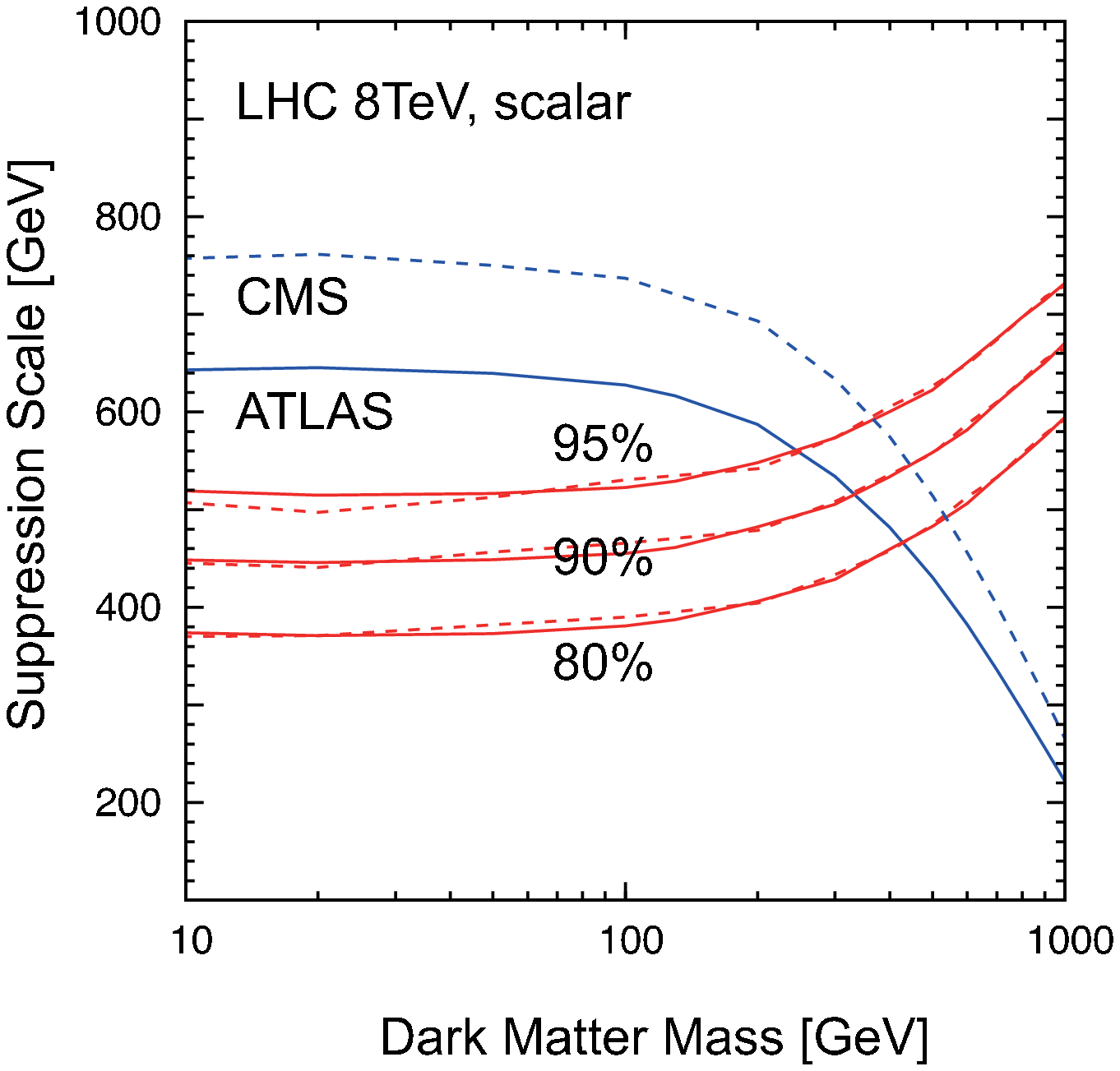}\hspace*{3mm}
 \includegraphics[width=7cm]{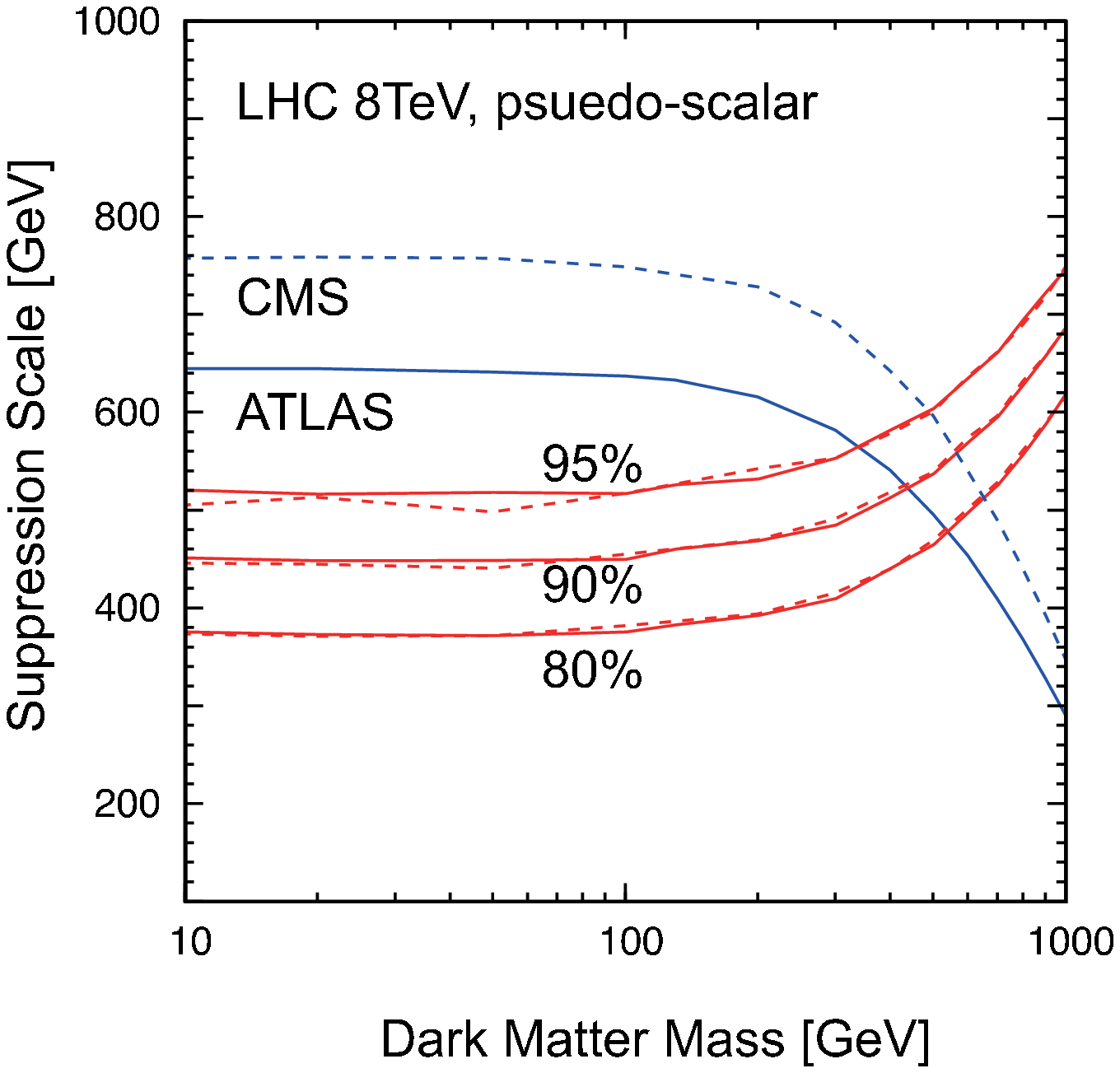}\vspace*{3mm}
 \includegraphics[width=7cm]{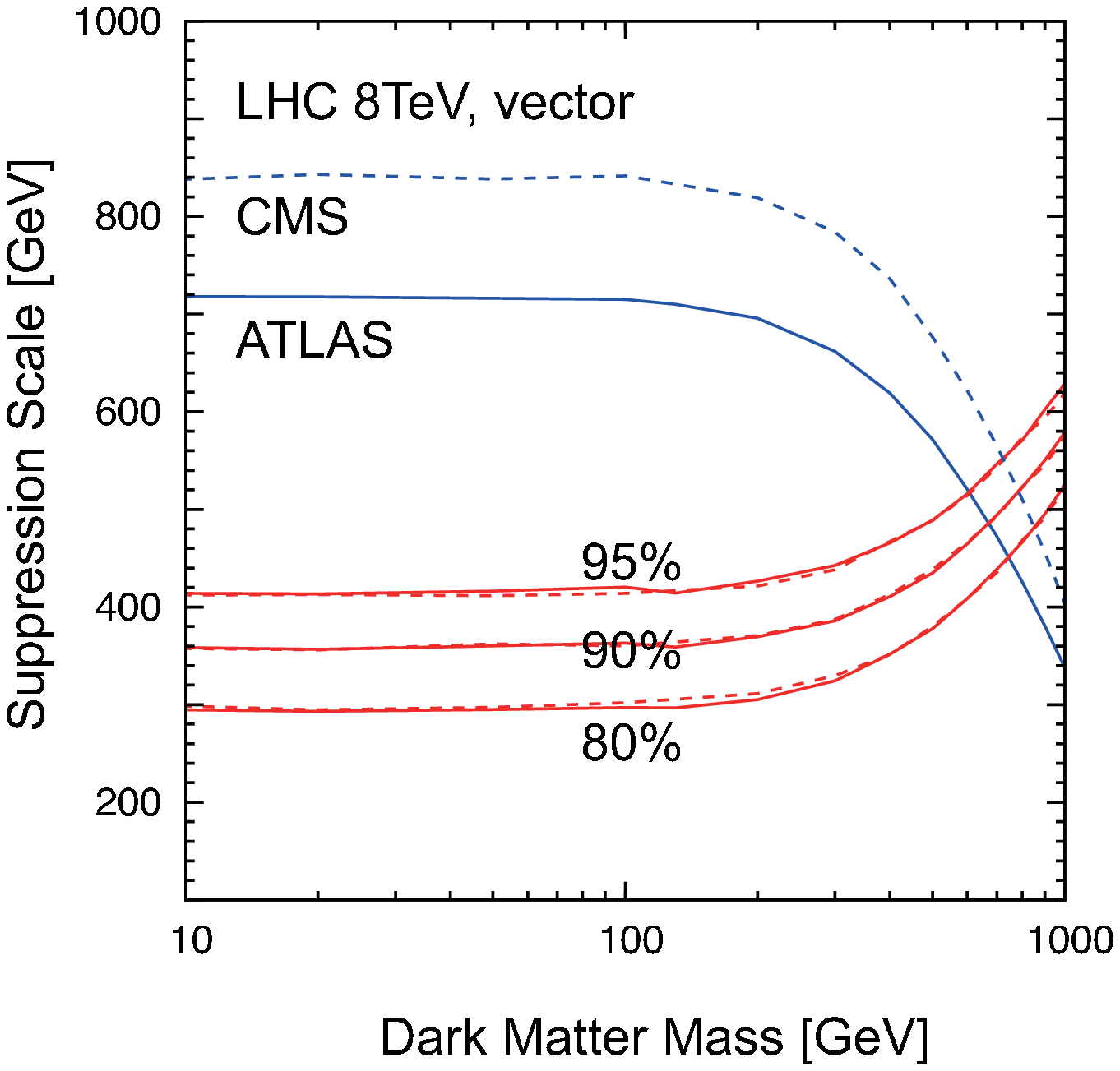}\hspace*{3mm}
 \includegraphics[width=7cm]{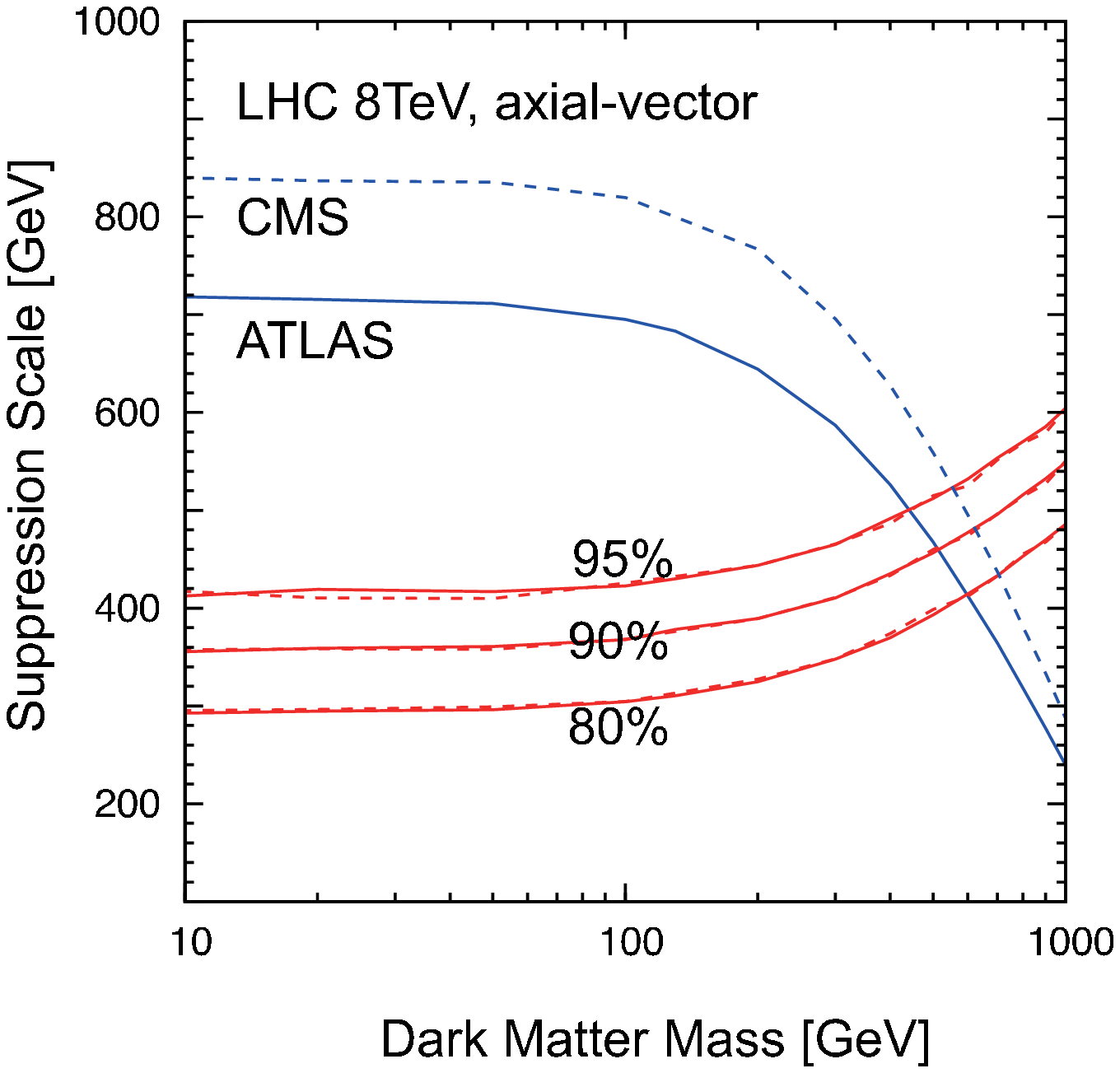}\vspace*{3mm}
 \includegraphics[width=7cm]{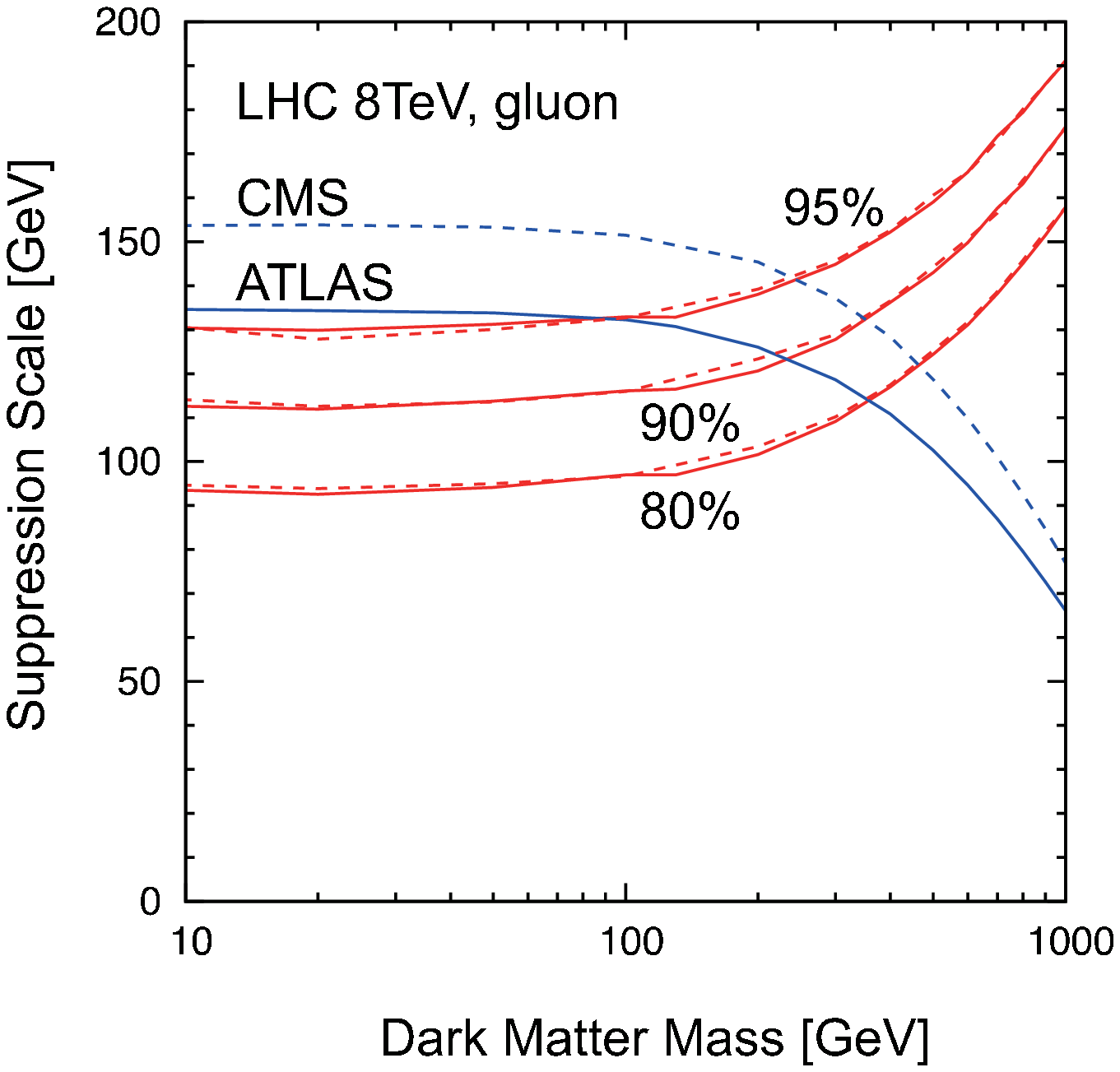}
 \caption{Unitarity bound (red lines) and monojet constraint (blue lines) on the suppression scale $\La$.
 The solid lines are obtained for ATLAS with $\sqrt{s}=8\TeV$ and $\mcl{L}=10.5\invfb$, while the dashed lines are for CMS with $\sqrt{s}=8\TeV$ and $\mcl{L}=19.5\invfb$.}
 \label{fig:LHC8TeV}
\end{figure}

In Ref.~\cite{ATLAS:2012zim}, the monojet signature is studied by ATLAS.
The energy and integrated luminosity are $\sqrt{s}=8\TeV$ and $\mcl{L} = 10.5\invfb$, respectively. 
The monojet signals are searched for with the following cuts.
At first, they are chosen by the preselection conditions:
\begin{itemize}
\item The missing transverse energy satisfies $E_T^{\,\rm miss} > 120\GeV$.
\item The leading jet is required to have $p_T > 120\GeV$ in $|\eta| < 2$.
\item At most two hard jets with $p_T > 30\GeV$ and $|\eta| < 4.5$ are allowed. 
\item The second-leading jet satisfies $\Delta\phi(j_2, E_T^{\,\rm miss}) > 0.5$.
\item The events are vetoed, if the events include reconstructed leptons. 
\end{itemize}
The trigger conditions are included in this preselection. 
Their efficiencies are sufficiently good. 
After the preselection, the dark matter signals are studied in the signal region SR3 of Ref.~\cite{ATLAS:2012zim}, which is defined to be:
\begin{itemize}
\item The leading jet and MET satisfy $p_T(j_1) > 350\GeV$ and $E_T^{\,\rm miss} > 350\GeV$.
\end{itemize}
Since no excess has been observed in the experimental data over the backgrounds, the dark matter production cross section, $\sigma \times A \times \epsilon$, is limited to be less than $0.05\pb$ at 95\% CL, where $A$ and $\epsilon$ are acceptance and efficiency, respectively.
In this letter, let us call this as the monojet limit.

For the analysis of the unitarity, the monojet events are generated by the Monte Carlo simulation.
The model file is constructed by {\tt FeynRules} v2.0.20~\cite{Alloul:2013bka}.
The events, $pp \to \chi\bar\chi j$, are generated at the leading order by {\sc MadGraph5} v2.1.0~\cite{Alwall:2011uj} with CTEQ6L1 PDFs~\cite{Pumplin:2002vw}. 
The renormalization scale of $\al_s$ is set by following the convention of {\sc MadGraph5}, particularly, the geometric mean of $(m^2_\text{DM}+p_T^2)^{1/2}$ for the gluon effective operator. 
The generated events are interfaced to {\tt Pythia} v6.426~\cite{deFavereau:2013fsa} for the parton shower and hadronization, and to {\tt Delphes} v3.0.12~\cite{Sjostrand:2006za} for the detector simulation. 
The default ATLAS card of {\tt Delphes} is used for the detector setup, but the jet radius parameter $R = 0.4$ is set for the anti-kt algorithm in accord with the ATLAS analysis. 
In order to validate this procedure, the monojet limit is calculated for the vector, axial-vector and gluon effective operators. 
We checked that our results agree very well with the ATLAS ones~\cite{ATLAS:2012zim}. 

The unitarity bound is imposed on the signal events that pass the ATLAS SR3 cut condition.
In Fig.~\ref{fig:LHC8TeV}, the bounds on the suppression scale $\La$ are shown by the red solid lines for each effective operator. 
For $m_{\rm DM} \lesssim 200\GeV$, the bound is insensitive to the dark matter mass.
It becomes stronger when the mass is larger, regardless of the phase space suppression, since larger $\bar s$ is required to satisfy the SR3 cut conditions.
In the figure, the numbers associated to the red lines mean that, among the total signal events that pass the SR3 cut condition, 95\%, 90\% and 80\% of the events satisfy the unitarity bound, respectively.
In other words, the signal events include 5\%, 10\% and 20\% of the events for which the effective Lagrangian approach breaks down. 
They should be calculated with UV completions.
Since we do not know what the UV physics is, we naively adopt a theoretical uncertainty to the number of signal events by 5\%, 10\% and 20\%, respectively, by assigning 100\% errors to the number of the unitarity-violating events.
The uncertainty is added to those already included in the ATLAS analysis. 

The unitarity bound is compared with the monojet limit. 
The monojet limit from ATLAS is displayed by the solid blue line in Fig.~\ref{fig:LHC8TeV}. 
It becomes weaker for heavier dark matter masses, since larger energy is required to produce dark matters.
If the unitarity bound is stronger than the monojet limit, the effective theory approach is not appropriate for the analysis. 

For the vector operator, the unitarity bound is satisfied in a wide dark matter mass range of the monojet limit. 
The ATLAS analysis includes about 10\% (5\%) events violating the unitarity condition for $m_{\rm DM} < 700\GeV$ ($600\GeV$).
According to Ref.~\cite{ATLAS:2012zim}, systematic uncertainties in the theoretical calculation comes from PDFs, renormalization/factorization scales and modelling of ISR/FSR jets, apart from the unitarity. 
They are typically estimated to be about 10\% for the vector operator with small dark matter masses, and become larger as the dark matter is heavier. 
Thus, the monojet limit is reliable for $m_{\rm DM} < 600$--$700\GeV$, at least, in terms of the above unitarity condition.

The axial-vector operator has a similar result as the vector one. 
The systematic uncertainties in the theoretical calculation are also expected to be similar to the vector case. 
If the theoretical uncertainty from the unitarity is required to be less than 10\% (5\%), the dark matter is constrained to be lighter than $m_{\rm DM} < 500\GeV$ ($400\GeV$) to use the monojet limit.
If the dark matter is heavier than it, the theoretical uncertainty becomes large.
On the other hand, the scalar and pseudo-scalar effective operators have stronger bounds from the unitarity condition compared to the monojet limit, as found in Fig.~\ref{fig:LHC8TeV}.
The monojet analyses pass the unitarity conditions in $m_{\rm DM} < 200$--$300\GeV$ for the scalar operator and in $m_{\rm DM} < 300$--$400\GeV$ for the pseudo-scalar operator, if the uncertainty from the unitarity is required to be less than 5--10\%.

The unitarity bound becomes much stronger against the monojet limit for the gluon operator. 
The former bound on the suppression scale is naively given by the c.m.~energy of the subprocess. 
As the dimension of the effective operator increases, the bound tends to be stronger, as noticed by comparing the gluon operator with the others in Sec.~\ref{sec:lagrangian}. 
On the other hand, the systematic uncertainties in the theoretical calculation are estimated to be large for the gluon initial state, according to Ref.~\cite{ATLAS:2012zim}. 
They are typically $\sim 30\%$ or larger for the gluon operator.
If the uncertainty from the unitarity violation is limited to be less than 20\%, the dark matter mass is bounded to be smaller than $300\GeV$ for the monojet limit.
As a result, the analysis of the ATLAS monojet limit is safe in terms of the unitarity bounds up to $m_{\rm DM} = \mcl{O}(100)\GeV$, depending on the effective operators.
For heavier dark matters, additional theoretical uncertainty should be assigned together with those already taken into account in the monojet limit.

\subsection{CMS at 8\,TeV}
\label{sec:CMS}

CMS analyzes the monojet signatures similarly with ATLAS. 
It provides a stronger limit than ATLAS, since CMS have analyzed more data.
In Ref.~\cite{CMS:rwa}, the integrated luminosity is $\mcl{L} = 19.5\invfb$ for $\sqrt{s}=8\TeV$. 
In order to study the monojet signals, the events are selected by the following preselection cuts:
\begin{itemize}
\item The event has MET with $E_T^{\,\rm miss} > 200\GeV$.
\item The leading jet satisfies $p_T(j_1) > 110\GeV$ in $|\eta(j_1)| < 2.4$.
\item Events with more than two jets with $p_T > 30\GeV$ and $|\eta| < 4.5$ are vetoed.
\item The second jet is allowed, provided that it satisfies $\Delta\phi(j_1, j_2) < 2.5$.
\item The event is discarded if it includes reconstructed leptons or taus. 
\end{itemize}
The trigger conditions are included in these cuts with sufficiently good efficiencies. 
The dark matter signals are analyzed in the signal region:
\begin{itemize}
\item MET satisfies $E_T^{\,\rm miss} > 400\GeV$.
\end{itemize}
Since no excess has been observed, the number of signals is limited to be less than 434 at 95\% CL. 
In this letter, the signal events are simulated in the same way as Sec.~\ref{sec:ATLAS}, but the default CMS card of {\tt Delphes} is used for the detector simulation with the jet radius $R = 0.5$ in accord with the CMS analysis.
We checked that the calculated monojet limits agree well with the CMS result~\cite{CMS:rwa}. 

In Fig.~\ref{fig:LHC8TeV}, the unitarity bounds and the monojet constraints are shown by the dashed lines.
The latter is stronger than the ATLAS result, since the analyzed luminosity is larger. 
On the other hand, the unitarity bound is almost the same as the ATLAS one with the SR3 cut conditions. 
In ATLAS, $p_T(j_1) > 350\GeV$ and $E_T^{\,\rm miss} > 350\GeV$ are imposed, whereas CMS uses $E_T^{\,\rm miss} > 400\GeV$.
Since the transverse momentum of the jet is equal to that of the $\chi\bar\chi$ pair at the parton level, the CMS cut condition is substantially close to that of ATLAS except for a small difference of the energy threshold. 
Although the unitarity bound depends on the cut condition, as will be discussed in Sec.~\ref{sec:LHC14TeV}, this difference is sufficiently small.
Therefore, the unitarity bound of CMS becomes almost the same as that of ATLAS.

If we compare the unitarity bound with the monojet limit, the conclusion is similar to the above analysis in Sec.~\ref{sec:ATLAS}.
The CMS monojet analyses are safe against the unitarity bound in the region $m_{\rm DM} < \mcl{O}(100)\GeV$, depending on the effective operators.
In detail, the upper bounds on the dark matter mass in these analyses become higher by about $100\GeV$ compared to ATLAS.

\subsection{Discussions}
\label{sec:discussion}

In this section, we have studied the unitarity bound on the dark matter monojet analyses at LHC. 
The unitarity conditions are calculated for $2\to2$ scatterings, whereas the the monojet processes \eqref{eq:process} are $2\to3$ processes.
Although it is possible to derive the unitarity condition for $2\to3$ scatterings, it turns out to be too weak to restrict the effective operators.
Instead, we focus on $2\to2$ subprocesses of the dark matter effective vertices.
For such vertices, one of the initial particles becomes off-shell. 
If the virtuality is sufficiently small, the unitarity conditions would be applicable.
In detail, monojet processes include three types of diagrams with respect to the effective vertex: $2\to2$, $1\to3$ and $2\to3$ scatterings. 
Here, $1\to3$ subprocesses mean diagrams of s-channel quark/gluon exchanges, e.g., $pp \to q^\ast \to \chi\bar\chi q$ for the vector effective operator, while $2\to2$ subprocesses represent t-channel diagrams.
At LHC with $\sqrt{s}=8\TeV$, $1\to3$ and $2\to3$ contributions amount to at most $\sim10\%$ of the signal events.
Further, for $2\to2$ subprocesses the virtuality of an initial particle is represented by the momentum transfer $Q^2$ of the off-shell particle. 
We checked that $Q^2$ is typically less than $\sim 10\%$ of $\bar s$ for the signal events that violate the unitarity bound. 
The virtuality becomes smaller for signals of heavier dark matters. 
Thus, the effect of the off-shellness is considered to be subdominant in the analysis of the unitarity.

So far, the unitarity bounds and the monojet limits have been discussed for each effective operator.
Even if there are multiple operators simultaneously, the conclusion does not change except for the quantitative details. 
It may be instructive how to apply the above analysis to general cases.
When there are multiple effective operators, they contribute to the signal events simultaneously. 
If a scattering amplitude of an operator interferes with others, the unitarity bound is derived from the helicity amplitude as in Sec.~\ref{sec:lagrangian}. 
The amplitude may not be represented by a single term in the partial wave expansion.
Then, the constraint is imposed on each term by Eq.~\eqref{EqUnitarity}.
On the other hand, if a process includes scattering amplitudes which do not interfere with each other, the unitarity bound is derived for each scattering rate, and the monojet analysis is tested against the bound in the same way as explored in this section.

\section{Prospects for LHC 14TeV}
\label{sec:LHC14TeV}

\begin{figure}[htb]
\begin{center}
 \includegraphics[width=7cm]{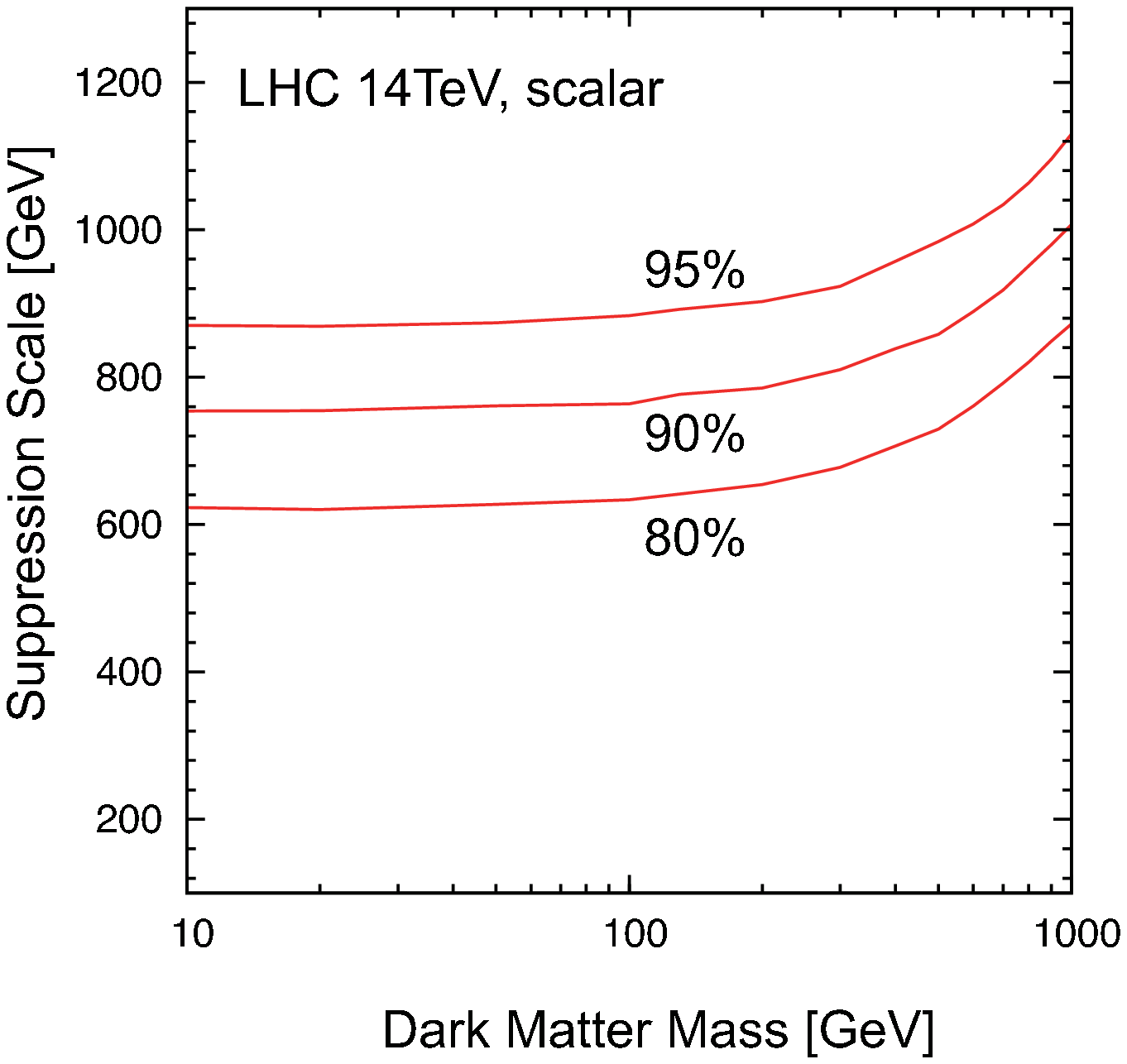}\hspace*{3mm}
 \includegraphics[width=7cm]{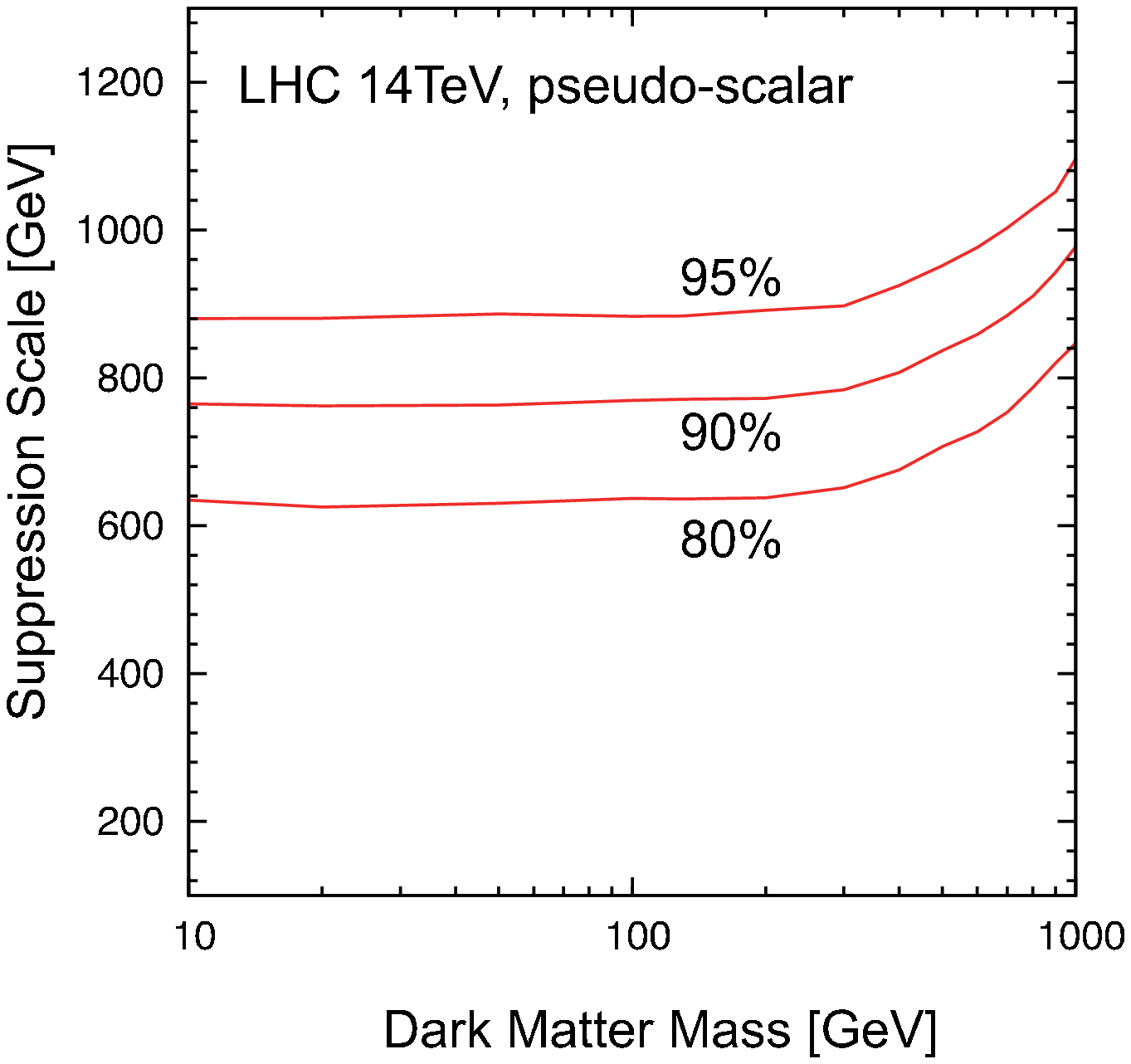}\vspace*{3mm}
 \includegraphics[width=7cm]{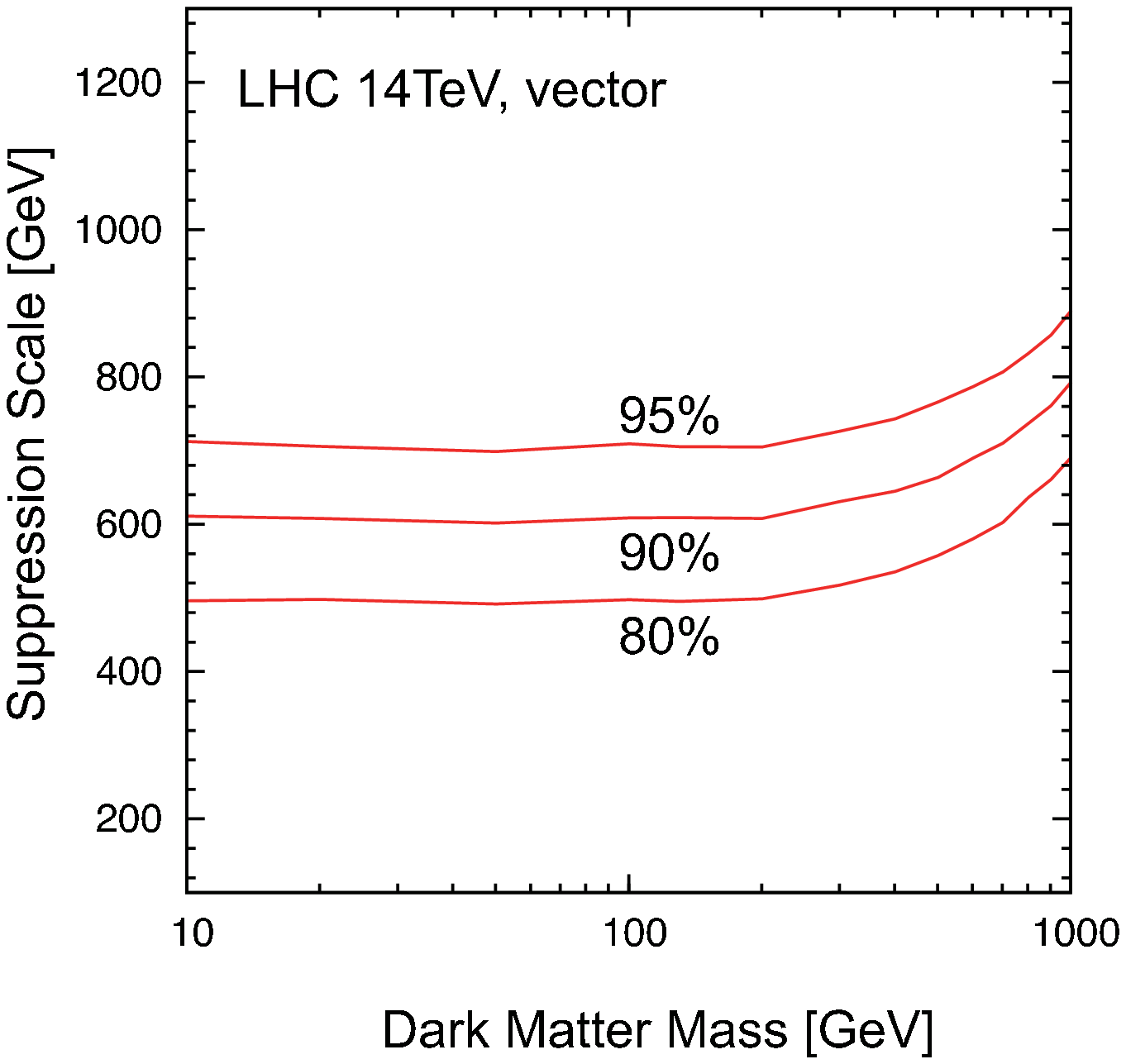}\hspace*{3mm}
 \includegraphics[width=7cm]{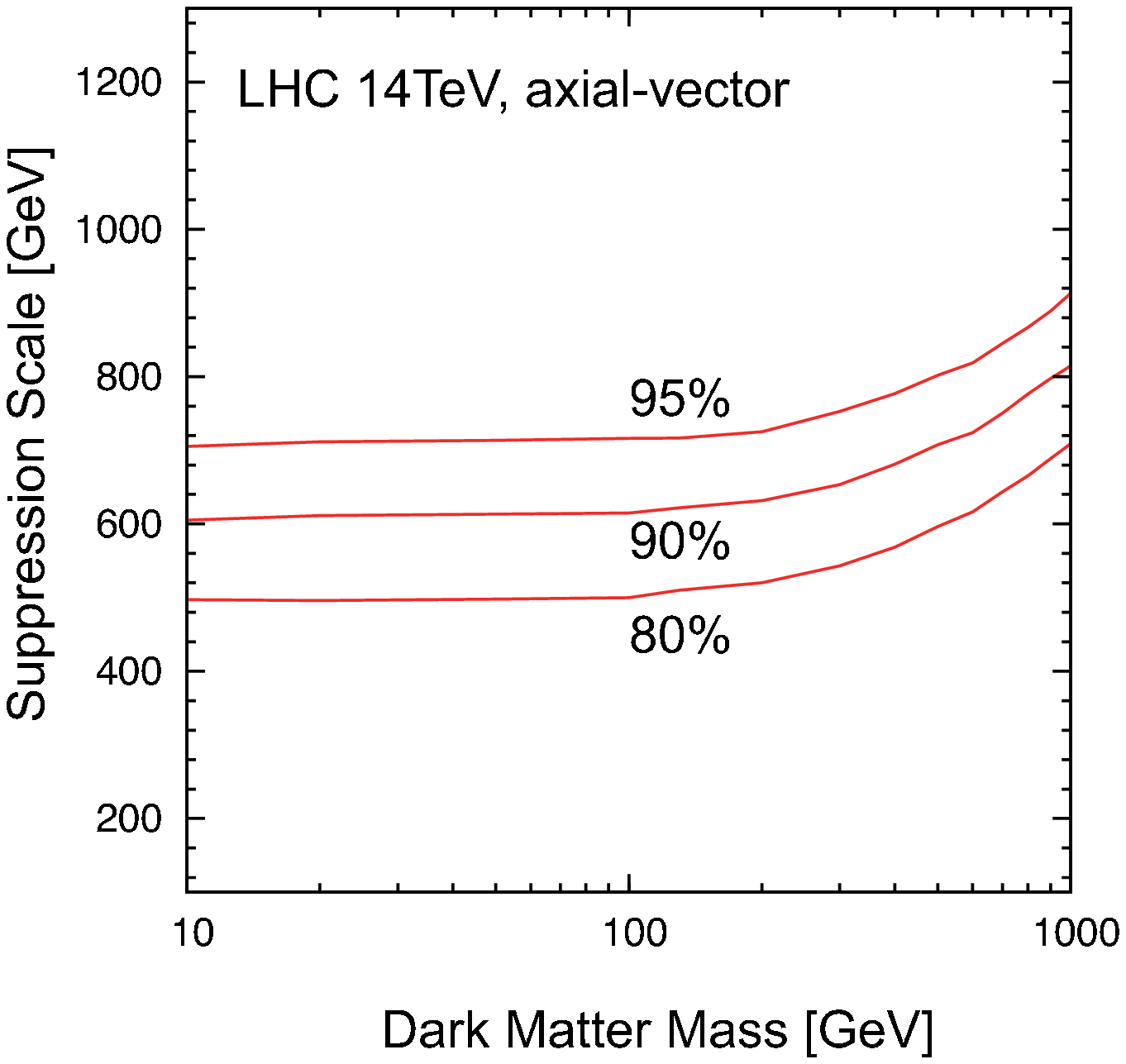}\vspace*{3mm}
 \includegraphics[width=7cm]{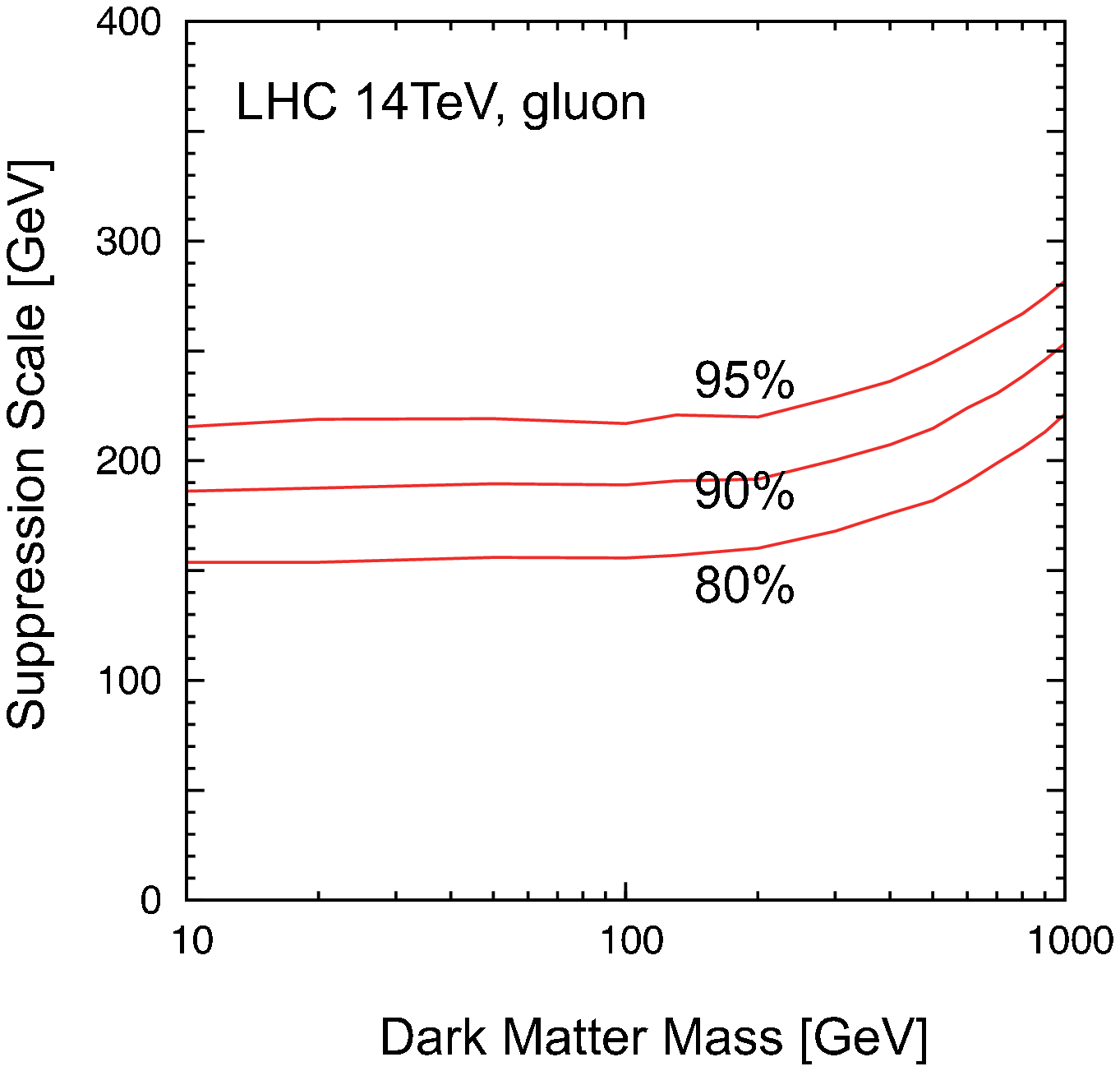}
\end{center}
 \caption{Unitarity bound at LHC $\sqrt{s} = 14\TeV$. The preselection is the same as ATLAS $8\TeV$, while the cut condition is set to be $p_T(j_1) > 500\GeV$ and $E_T^{\,\rm miss} > 500\GeV$.}
 \label{fig:LHC14TeV}
\end{figure}

\begin{figure}[htb]
\begin{center}
 \includegraphics[width=7cm]{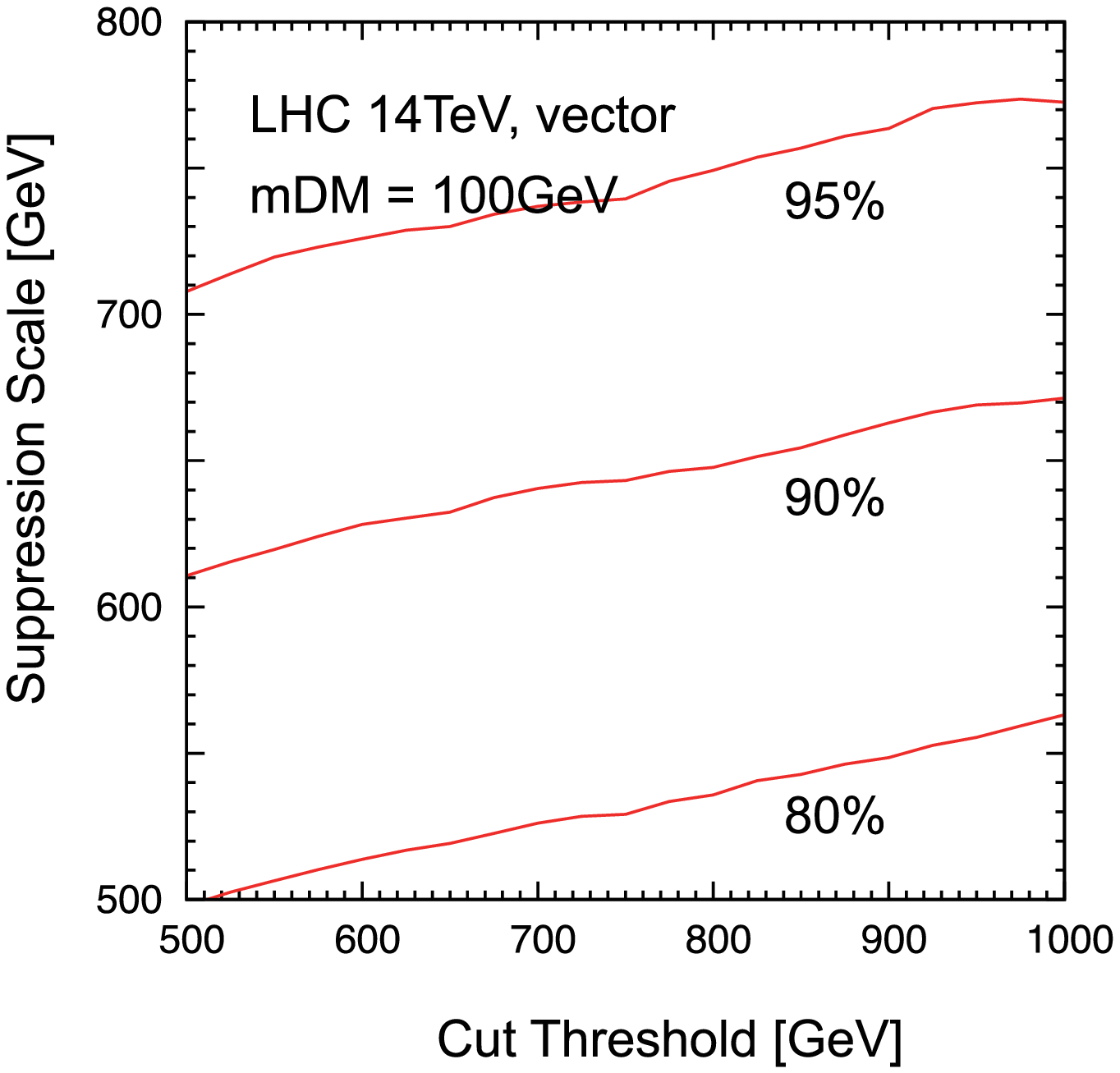}\hspace*{5mm}
 \includegraphics[width=7cm]{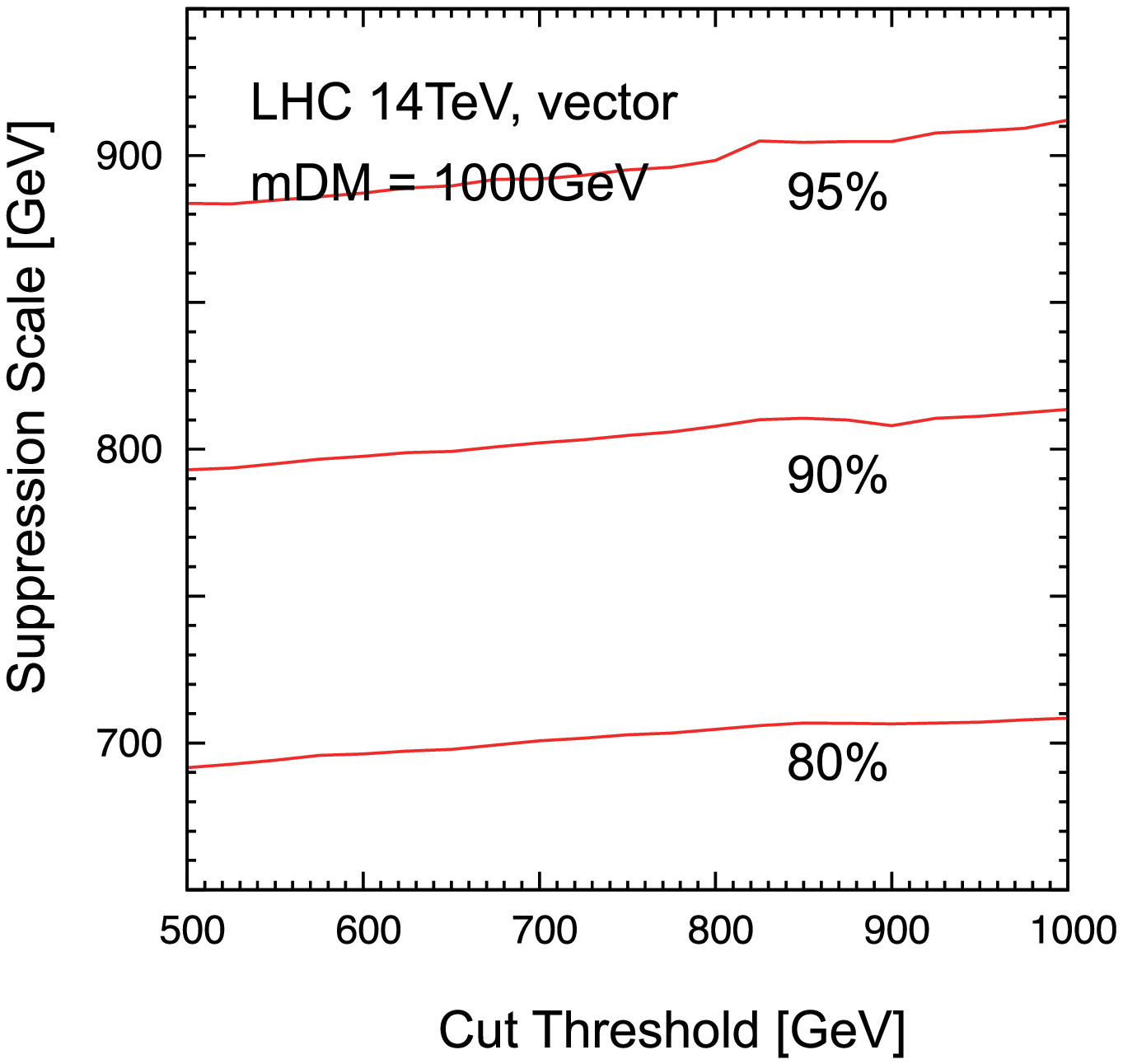}
\end{center}
 \caption{
   Cut dependences of the unitarity bound for the vector operator.
   The horizontal axes are the cut threshold of $p_T(j_1)$ and $E_T^{\,\rm miss}$.
   For the signal events, the transverse momentum of the leading jet and MET are required to be larger than the threshold. 
	 The vector effective operator is analyzed with $m_{\rm DM} = 100$ (left) and $1000\GeV$ (right).
	 The vertical axes are the suppression scale of the operator, $\La_V$.
}
 \label{fig:cut}
\end{figure}

As the collision energy increases, the unitarity bound becomes stronger, since the c.m.~energy of the subprocess tends to be larger.
The bound is estimated for LHC at $\sqrt{s} = 14\TeV$. 
In order to simulate the bound, the same event preselection as ATLAS $8\TeV$ is imposed.
The signal region, $p_T(j_1) > 500\GeV$ and $E_T^{\,\rm miss} > 500\GeV$, is set as a test case. 
The result is shown in Fig.~\ref{fig:LHC14TeV}.
It is found that the behaviour of the unitarity bound is the same as that obtained at $\sqrt{s} = 8\TeV$, whereas the bound on the suppression scale increases by about $300\GeV$ for the scalar, pseudo-scalar, vector and axial-vector operators. 
And, it becomes stronger by about $100\GeV$ for the gluon operator. 

The unitarity bound depends on cut conditions to select signals. 
If stronger cut on $p_T(j_1)$ or $E_T^{\,\rm miss}$ is imposed, the signal events are likely to have larger c.m.~energy, $\bar s$, in the subprocess $ii \to \chi\bar\chi$. 
Thus, the unitarity bound becomes stronger when the cut threshold is tightened. 
In Fig.~\ref{fig:cut}, the cut dependence is displayed for dark matter masses, $m_{\rm DM} = 100$ and $1000\GeV$. 
Here, the preselection of the signal events is the same as ATLAS $8\TeV$ in Sec.~\ref{sec:ATLAS}, while the signal region is defined such that the events satisfy $p_T(j_1)$ and $E_T^{\,\rm miss}$ larger than the cut threshold momentum/energy (the horizontal axis). 
It is found that the unitarity bound on the suppression scale increases by $50$--$100\GeV$ if the cut threshold is varied from $500$--$1000\GeV$.
Also, the signals with lighter dark matter are more sensitive to the cut conditions than those with heavier one, though the difference is small. 
Backgrounds are expected to be large in the next phase of LHC.
Tight cut conditions can help to suppress them in the analysis of the monojet signature.
Then, the perturbative unitarity bound becomes more significant.

In the figure, the vector effective operator is analyzed to study the cut dependence of the unitarity bound. 
The conclusion does not change for the other effective operators.

\section{Conclusion}
\label{sec:conclusion}

We have studied the effect of the unitarity condition in the monojet process, where the production of the dark matter is described by the low-energy effective theory.
The theoretical calculation of the dark matter signal is, at least, required to satisfy the perturbative unitarity condition. 
In this letter, the condition is discussed in detail.
It was found that the LHC analysis based on the effective theory is not appropriate when the dark matter is heavier than several hundreds GeV.
UV physics is required for the analysis of the monojet signal in larger dark-matter mass regions. 
The unitarity bound is independent of the UV physics and cannot be avoided, for instance, by tuning the mediator between the dark matter and the SM particles.

In this letter, we focused on the monojet signature of the Dirac dark matter.
It is straightforward to apply this analysis to the other setup, such as a scalar dark matter or  the mono-Z, W signatures. 
These studies will be explored in the near future.

\section*{Acknowledgement}
This work was supported by JSPS KAKENHI Grant No.~23740172 (M.E.).
The work of Y.Y. has been supported in part by the Ministry of Economy and Competitiveness (MINECO), grant FPA2010-17915, and by the Junta de Andaluc\'ia, grants FQM 101 and FQM 6552.

\appendix
\section{Wave functions in helicity basis}
\label{sec:helicity}
\subsection{Spin 1/2}

In the Dirac representation, spinors in helicity eigenstate which propagate to the $(\theta,\phi)$ direction with energy $E$ and helicity $\la =\pm 1/2$ are represented as
\begin{align}
 u(\vec{p},\la) =& \begin{pmatrix}
  \sqrt{E+m} \ch_\la (\hat{\vec{p}}) \\
  2\la \sqrt{E-m} \ch_\la (\hat{\vec{p}})
 \end{pmatrix},\\
 v(\vec{p},\la) =& \begin{pmatrix}
  \sqrt{E-m} \ch_{-\la} (\hat{\vec{p}}) \\
  -2\la \sqrt{E+m} \ch_{-\la} (\hat{\vec{p}})
 \end{pmatrix},
\end{align}
where $m$ is a fermion mass, and the $\chi$ spinor is given by
\begin{align}
 \ch_{1/2} (\hat{\vec{p}}) = \begin{pmatrix}
  \cos \frac{\th}{2} \\ e^{i\ph} \sin \frac{\th}{2}
 \end{pmatrix}, \qquad
 \ch_{-1/2} (\hat{\vec{p}}) = \begin{pmatrix}
  -e^{-i\ph} \sin \frac{\th}{2} \\ \cos \frac{\th}{2}
 \end{pmatrix}.
 \label{eq:spin}
\end{align}
Spinors in the chiral representation are obtained by rotating the above spinors with the matrix,
\begin{align}
 M = \frac{1}{\sqrt{2}} \begin{pmatrix} 1 &-1 \\ 1 & 1 \end{pmatrix}.
\end{align}
Explicitly, they are shown as
\begin{align}
 u(\vec{p}, \la) =& \frac{1}{\sqrt{2}} \begin{pmatrix}
   \sqrt{E+m}-2\la \sqrt{E-m} \\ \sqrt{E+m}+2\la \sqrt{E-m}
 \end{pmatrix} \ch_\la ,\\
 v(\vec{p}, \la) =& \frac{1}{\sqrt{2}} \begin{pmatrix}
   \sqrt{E-m} +2\la \sqrt{E+m} \\ \sqrt{E-m} -2\la \sqrt{E+m}
 \end{pmatrix} \ch_{-\la} .
\end{align}
The $\chi$ spinor is the same as Eq.~\eqref{eq:spin}.
In this letter, we follow the Jacob-Wick convention \cite{Jacob:1959at} to represent helicity amplitudes. 
In $2 \to 2$ scattering amplitudes, the second spinor, which propagates opposite to the first one in the c.m.~frame, has an extra phase. See e.g.~Ref.~\cite{Haber:1994pe} for details. 

\subsection{Spin 1}
Polarization vectors of massless vector fields propagating to the $(\th ,\ph)$ direction are represented as
\begin{align}
 \ep_\pm^\mu =
 \frac{1}{\sqrt{2}} 
 e^{\pm i\ph}
 (0, \mp \cos\th \cos\ph +i\sin\ph ,\mp \cos\th \sin\ph -i\cos\ph ,\pm \sin\th ).
\end{align}
The overall phase is defined in accord with the Jacob-Wick convention~\cite{Jacob:1959at}. 
Also, the four momentum is shown as 
\begin{align}
 p = (p, p \sin\th \cos\ph ,p \sin\th \sin\ph ,p \cos\th).
\end{align}

\subsection{Wigner $d$-function}

The Wigner $d$-function is characterized by the representation of the rotation algebra.
Instead of showing a general result, let us list the explicit functions which are used in this letter:
\begin{align}
  & d^0_{0,0}(\th) = 1, \qquad
  d^1_{0,0}(\th) = \cos \th, \\ &
  d^1_{1,1}(\th) = d^1_{-1,-1}(\th) = \cos^2 \frac{\th}{2}, \qquad
  d^1_{1,-1}(\th) = d^1_{-1,1}(\th) = \sin^2 \frac{\th}{2}, \\ &
  d^1_{1,0}(\th) = -d^1_{-1,0}(\th) = -d^1_{0,1}(\th) = d^1_{0,-1}(\th).
  = -\frac{\sin \th}{\sqrt{2}},
\end{align}


\end{document}